\documentclass[showpacs,preprintnumbers,amsmath,amssymb]{revtex4}

\usepackage{graphicx}
\usepackage{dcolumn}
\usepackage{bm}

\begin{document}

\preprint{Physica A 371, 627(2006)}

\title{Optimized annealing of traveling salesman problem from the $n$th-nearest-neighbor distribution}

\author{Yong Chen\footnote[1]{Corresponding author. Email: ychen@lzu.edu.cn} and Pan Zhang\footnote[2]{Email: july.lzu@gmail.com}}

\address {Institute of Theoretical Physics, Lanzhou University, Lanzhou $730000$, China}

\date{\today}

\begin{abstract}
We report a new statistical general property in traveling salesman problem, that the $n$th-nearest-neighbor distribution of optimal tours verifies with very high accuracy an exponential decay as a function of the order of neighbor $n$. With defining the energy function as the deviation $\lambda$ from this exponential decay, which is different to the tour length $d$ in normal annealing processes, we propose a distinct highly optimized annealing scheme which is performed in $\lambda$-space and $d$-space by turns. The simulation results of some standard traveling salesman problems in TSPLIB$95$ are presented. It is shown that our annealing recipe is superior to the canonical simulated annealing.
\end{abstract}


\maketitle

The traveling salesman problem (TSP) is stated for the shortest closed tour for a traveling salesman who must visit each of $N$ cities in turn \cite{tsp1,tsp2,tsp3}. The number of candidate set is $N!/2N$. So it is very difficult to find an efficient algorithm for large $N$ because it is bounded by a polynomial function with the problem size. For example, Padberg and Rinaldi obtained the optimal solution, i.e., the shortest path, for $532$ U.S. cities after $6$ hours of calculation with the use of the supercomputer Cyber $205$ in $1987$ \cite{1}. And, the TSP is a classic famous nondeterministic polynomial problem (NP-complete) and a good testing ground for optimization methods. Because the exact solutions are almost impossible to obtain, the realistic and valuable aim is to seek for a near-optimal solutions.

In recent several decades, TSP has attracted great attention of many physicists, and a lot of algorithms with physical insight had been proposed. In $1982$, Hopfield and Tank used neural networks to find the approximate solution \cite{nn1}, and since then, simulated annealing (SA) \cite{sa1,sa2,combine}, hierarchical constructioncite \cite{hy1}, real space renormalization \cite{re1,re2,re3}, diffusion process \cite{diff}, etc. had been introduced to challenge TSP.

Among these algorithms from statistical physics, the most general powerful solution is simulated annealing scheme \cite{sa1,sa2}. The classical simulated annealing (CSA) was proposed by Kirkpatrick {\it et al.} \cite{sa1} that it extended the Metropolis procedure for equilibrium Boltzmann-Gibbs statistics. Moreover, it was shown that the system will end in a global minimum if the temperature decreases as the inverse logarithm of time \cite{gemann1984}. This annealing process allows the exploration of the configuration space avoiding trapping into the local minima of energy function. However, in fact, since the configuration space is bumpiness, CSA has to use higher initial temperature and longer annealing time to escape from the energy valleys. So the open question is how to get away from the energy valleys more efficiently. Then some complex methods had been proposed. Szu and Hartley proposed the so called fast simulated annealing (FSA) that the system can jump around the energy landscape due to a Cauchy-Lorentz visiting distribution instead of the Gaussian in CSA \cite{Szu1987}. The subsequent generalized version presented by Tsallis and Stariolo from Tsallis statistics and it was applied to TSP for faster stochastic method of SA \cite{Tsallis1994,Penna1995}. Another effective advanced recipe is quantum annealing (QA) \cite{qa1,qa2} where the quantum fluctuations instead of thermal fluctuations to tunnel through energy valleys.

However, in this paper, different than the traditional solutions to optimize the escape from the local energy minima, we focus our attention on how to construct a new flatter configuration space where the annealing process should be easier, faster, and more efficient to reach the final near-optimal or optimal solutions.

Our recipe of TSP can be simply stated as the following description. Noted that there are a large number of optimal solutions of TSPs in TSPLIB$95$ \cite{tsplib}. Through summarizing the statistical property of the $n$th-nearest-neighbor distributions (NNDs) of optimal tours of TSPs in TSPLIB$95$, it is found that the distributions decay exponentially with very high accuracy as a function of the order of neighbor $n$. And, we defined a new energy function and constructed a new configuration space under the name of $\lambda$-space, where $\lambda$ describes the deviation between the tour's neighbor distribution and the general decayed exponentially statistical property. A new simulated annealing is accomplished in $\lambda$-space and, further, is executed in $\lambda$-space and $d$-space by turns. This solution is considerably faster than the traditional SA in solving the TSP. It is indeed so, as shall be shown in the following text.

One considers a set of $N$ cities with intercity distance $d_{ij}$ between the $i$th city and $j$th city. In this work, the problems are limited to the symmetric TSPs, $d_{ij}=d_{ji}$. Especially, the $d_{ij}=\infty$ if there is no link between the $i$th node and $j$th node. The order of neighbor $n$ is defined by the nearest neighbor sequences. For example, the nearest neighbor is corresponding to $n=1$. $n=2$ means the edge $d_{ij}$ is the second shortest edge among all edges connecting with city $j$, or city $i$ is the second nearest neighbor of city $j$. The rest, $n=3, 4,\ldots, N-1$, may be deduced by analogy. For every possible tour of TSP, the $n$th-NND $r(n)$ should be defined by
\begin{equation}
r(n) = \frac{s(n)}{2N},\qquad n = 1,2,3,\ldots,N-1,
\label{eq01}
\end{equation}
where $s(n)$ is total number of the $n$th-nearest-neighbor for all links in this tour. Clearly, the $n$th-NND $r(n) \in [0,1]$ and $\sum_{n=1}^{\infty}{r(n) = 1}$. Idealistically, the best economical tour routine is that every link between two cities is shortest, or it only exists the nearest neighbor, $r(1) = 1$. Apparently, it is impossible found this extreme condition in realistic TSPs.

\begin{figure}
\centerline{\resizebox{11cm}{!}{\includegraphics{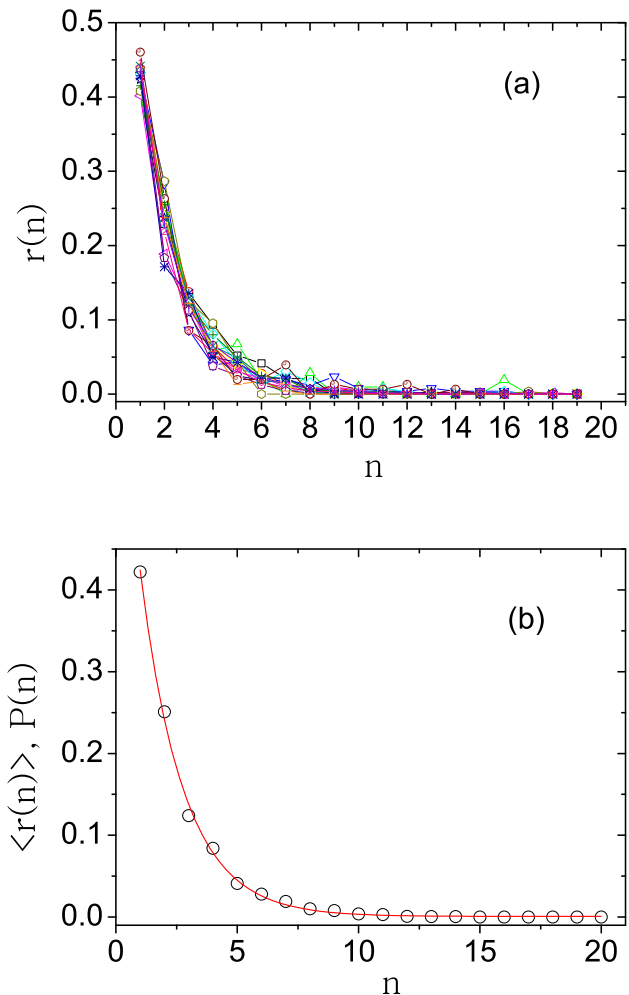}}}
\caption{(Color online) (a) the $n$th-NND $r(n)$ of optimal tours from att$48$ problem to rl$5934$ problem in TSPLIB$95$ vs the neighbors' order $n$. (b) The average $\langle r(n)\rangle$ (circles) and the fitting result $P(n)$ (solid line) as functions of $n$. A perfect exponential decays of $P(n)$ vs $n$ is shown with $P(n) = 0.74475 \times e^{-n/1.76985} + 0.000771$.}
\label{fig1}
\end{figure}

In Fig. \ref{fig1}(a), we presented the $n$th-NNDs $r(n)$ of optimal tours of many problems in TSPLIB$95$. It appears that all $r(n)$ can be qualitatively regarded as an exponential decay as a function of the order of neighbor $n$. And, it is should be pointed out that this qualitative exponential decay was confirmed by almost all TSPs in TSPLIB$95$. Then, the average $n$th-NND $\langle r(n) \rangle$ is given in Fig. \ref{fig1}(b). As a result of nonlinear fitting, the general empirical rule of the $n$th-NND is represented by
\begin{equation}
P(n) = A \times e^{-n/B} + C, \qquad n = 1,2,3,\ldots,N-1.
\label{eq02}
\end{equation}
Here $A = 0.74475 \pm 0.01376$, $B = 1.76985 \pm 0.03785$, and $C = 0.000771 \pm 0.00129$. Noted that $P(n)$ is introduced to denote the general empirical property, but $\langle r(n) \rangle$ is the statistical result. From the above considering, the $n$th-NND of optimal tour for any TSPs should be consistent with Eq. (\ref{eq02}). On the contrary, it is only stated that a tour with similar distribution of Eq. (\ref{eq02}) may be not optimal, but it must be an approximate tour.

\begin{figure}
\centerline{\resizebox{11cm}{!}{\includegraphics{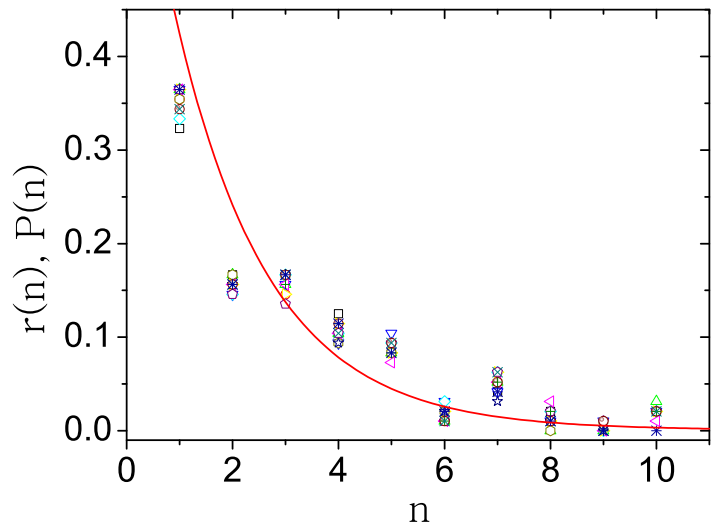}}}
\caption{(Color online) The neighbor distributions $r(n)$ of some approximate tours for att$48$ problem which have $10\%\sim20\%$ longer than the optimal tour length. The solid line is the general empirical neighbor distribution $P(n)$ of optimal tours from Eq. (\ref{eq02}).}
\label{fig2}
\end{figure}

An example of the comparison between $r(n)$ of the near-optimal tours and $P(n)$ for the att$48$ problem is given in Fig. \ref{fig2}. It is clear that there exists distinct deviation of the near-optimal $r(n)$ from the corresponding statistical $P(n)$. The difference of the $n$th-NND between any tours and the general empirical optimal tour Eq. (\ref{eq01}) is defined as the following equation,
\begin{equation}
\lambda = \sum_{i=1}^{n}{(s_{n}-\bar{s}_{n})\times{n}},
\label{eq03}
\end{equation}
where $s_{n}$ is the sum of number of the $n$th neighbor in this tour, and $\bar{s}_{n} = P(n)\times 2N$ is the empirical sum for the corresponding optimal tour from Eq. (\ref{eq01}) and Eq. (\ref{eq02}). Another important parameter is the tour length $d$ which is used as the energy function in most of simulated annealing algorithms, such as CSA, QA, etc..  Our annealing scheme is attempt to execute the Monte Carlo (MC) process to make the $n$th-NND $r(n)$ of any initial tour getting close to the empirical distribution of optimal tour $P(n)$. We will show the detail procedures in the following text.

The first crucial problem is how to generate a tour which should have smaller $\lambda$. The possible candidate solutions as we know are, tour construction algorithms, tour improving algorithms, and composite algorithms. The tour construction algorithms gradually build a tour by joining a new city at each step. The tour improvement algorithms improve upon a tour by performing various exchanges. And, the composite algorithms is a combination of the above two algorithms. Among these methods, tour improving algorithm is powerful and easy to implement. A simple example is the so-called $2$-opt algorithm. For a given tour, the new tour with shorter length is constructed by replacing two links by another two links, and the following procedures go on this cooperation until no more improvement are possible. In simulated annealing, usually, the tour improving algorithm is embedded in Metropolis MC process to generate new tours at environment temperature.

In our algorithm, the similar mechanism is used to optimize a given tour by metropolise MC process. We first generate a random tour and set up a temperature which will be reduced very slowly. For every step, a new tour is reconstructed from the old one by $3$-exchange algorithm \cite{lin1965}. If $\Delta\lambda < 0$, where $\Delta\lambda = \lambda_{new}-\lambda_{old}$, or $e^{-\lambda/t}$ greater than a random number between $0$ and $1$, the new tour is accepted. Repeated applications of this MC step until $\lambda$ smaller than $\lambda_{c}$, where $\lambda_{c}$ means the acceptable difference between the neighbor distribution of the final tour and the empirical optimal $P(n)$ from Eq. (\ref{eq02}). Since it is impossible to offer the exact value of $\lambda_{c}$ for each TSP, we usually keep running the annealing step until the tour length $d$ cannot get smaller.

Note that both our recipe and classical simulated annealing are belonged to the annealing scheme. There are the same MC steps, but our energy function is $\lambda$ different from the tour length $d$ in classical simulated annealing. Moreover, the corresponding configuration space is absolutely different. We label our annealing scheme as $\lambda$-space annealing, and rename the canonical simulated annealing scheme as $d$-space annealing.

\begin{figure}
\centerline{\resizebox{11cm}{!}{\includegraphics{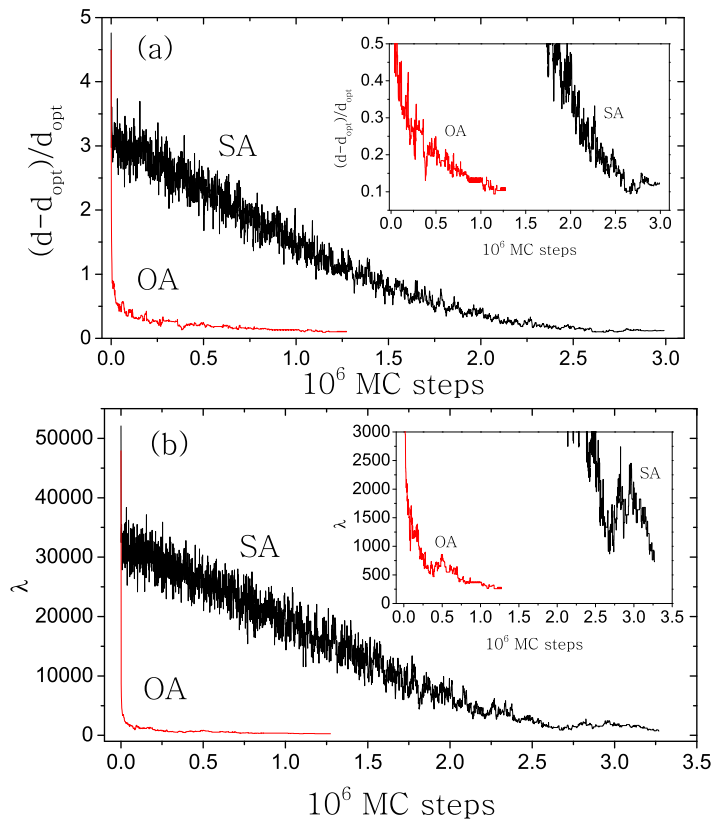}}}
\caption{(Color online) The detailed annealing processes in $d$-space annealing (SA) and $\lambda$-space (OA) for eil101 problem in TSPLIB$95$. The insets are the annealing details near optimal tour.}
\label{fig3}
\end{figure}

To measure the performances of the two distinct annealing strategies, we chose a standard TSP sample, eil$101$ problem, from the TSPLIB$95$ \cite{tsplib}. It is a $N=101$ cites problem constructed by Christofids and Eilon whose optimal tour length $d_{opt} = 629$. Both $\lambda$-space annealing and $d$-space annealing were applied to seek for the near optimal tours. Again, noted that both annealing schemes begin with a random tour and make use of the same $3$-exchange algorithm. In Fig. \ref{fig3}, we present the detail annealing processes in $d$-space and $\lambda$-space. The evolutions of tour length are described by $\Delta d =\frac{d-d_{opt}}{d_{opt}} \times 100\%$. As it turns out, the final near optimal tour for $\lambda$-space annealing approximately coincides with $d$-space annealing (see Fig. \ref{fig3}a and \ref{fig3}b). But the $\lambda$-space annealing is obviously faster than $d$-space annealing and noted that the initial temperature in $\lambda$-space annealing is far lower than in $d$-space annealing. In other words, $\lambda$-space annealing work at small initial temperature as well as at high initial temperature to make $\lambda$ of the tour smaller than the criterion $\lambda_{c}$. In fact, even only $3$-exchange technique can obtain a nice result, but it was highly dependent on the initial tours.

It is well known that the initial configuration with small temperature often cause trapping into local minimum too early. From Fig. \ref{fig3}, if someone uses $d$-space annealing to find a near optimal tour with same $\Delta d$ resulted from $\lambda$-space annealing, it should be paid out the higher initial temperature and the longer annealing time. So we conjecture that the configuration space in $\lambda$-space annealing is smoother than that in $d$-space annealing, or the fluctuation in $d$-space annealing is stronger than that in $\lambda$-space annealing. As a result, it is not necessary that one use the higher initial temperature to jump through energy valleys in $\lambda$-space annealing.

However, it is very important to mention that $d$-space annealing can increase accuracy at the cost of higher initial temperature and longer annealing time, but $\lambda$-space annealing lose this property because it almost produces the similar effect on annealing processes at higher or lower initial temperature for the more smooth $\lambda$-space. It was concluded that $\lambda$-space annealing can present approximate tours in shorter time, but very difficult to improve its accuracy.

In fact, the final tours obtained by $\lambda$-space annealing is really different than the tours obtained by $d$-space annealing though both $\Delta d$ are looked like the very similar value about $10 \%$ (see Fig. \ref{fig3}a). Taking the eil$101$ problem as example, the $\lambda$ of the approximate tour in $d$-space annealing is $730$ and the $\lambda$ in $\lambda$-space annealing is $260$ (several other examples presented in Table \ref{table-1}). The results in $\lambda$-space annealing have properties of small $d$ and small $\lambda$. It means that $\lambda$-space annealing can making $d$ of the tours close to the optimal tour as well as making the neighbor distributions $r(n)$ of tours close to $P(n)$. But in $d$-space annealing, the final approximate tours are restricted by the certain initial temperatures and their neighbor distributions are far from the distribution of optimal tour $P(n)$. Consequently, at a certain temperature, the annealing processes in $d$-space is easy to be trapped into local minimum and it loses the potential to find better tours in normal annealing schemes.

From the above statements, in $d$-space annealing, the tours usually have small $d$ and large $\lambda$ at local minimum. Similarly, in $\lambda$-space annealing, the tours normally have small $d$ and small $\lambda$, but they may be not in the local minimum of $d$-space at certain temperatures. It suggests that one can design a combinational optimized simulated annealing in both $d$-space and $\lambda$-space. As an example, it should be has higher potential to capture better tours when we take the tour obtained from $\lambda$-space annealing as the initial tour in $d$-space annealing at certain temperature. Theoretically, next, the annealing process can be moved onto $\lambda$-space. Repeated this annealing process back and forth until the global optimal tour was found. Simulated annealing in $\lambda$-space and $d$-space by turns is the point of our so-called optimized annealing scheme in this work. In fact, since the neighbor distribution of optimal tours $P(n)$ is just a statistical rule, we perform a simple case of optimized annealing under the name of $(\lambda +d)$-space annealing, which is constructed by $\lambda$-space annealing and $d$-space annealing only one time in this paper.

\begin{table*}[t]
\caption{Annealing results in $d$-space, $\lambda$-space, and $(\lambda +d)$-space for some problems in TSPLIB$95$.}
\begin{ruledtabular}
\begin{tabular}{c|ccc|ccc|ccc}

{} & {} &$d$-space& {} & {} &$\lambda$-space& {} & {} &$(\lambda +d)$-space & {} \\
\hline
Problems & $\frac{d-d_{opt}}{d_{opt}}$ & $\lambda$ & MC steps & $\frac{d-d_{opt}}{d_{opt}}$ & $\lambda$ & MC steps & $\frac{d-d_{opt}}{d_{opt}}$ & $\lambda$ & MC steps \\

\hline
Att$48$ & $5.1\%$ & $230$ & $3.0\times10^{6}$ & $5.0\%$ & $30$ & $0.8\times10^{6}$ & $0.1\%$ & $169$ & $0.2\times10^{6}$ \\
Pr$76$ & $5.1\%$ & $831$ & $3.6\times10^{6}$ & $6.8\%$ & $202$ & $2.2\times10^{6}$ & $3.4\%$ & $227$ & $0.1\times10^{6}$ \\
eil$101$ & $12.4\%$ & $730$ & $3.3\times10^{6}$ & $10.3\%$ & $260$ & $1.2\times10^{6}$ & $5.7\%$ & $351$ & $0.4\times10^{6}$ \\
Pr$107$ & $7.3\%$ & 800 & $2.3\times10^{6}$ & $7.1\%$ & 272 & $2.0\times10^{6}$ & $4.7\%$ & $373$ & $0.2\times10^{6}$\\
Pr$124$ & $8.9\%$ & 3138 & $3.0\times10^{6}$ & $16.2\%$ & 359 & $0.1\times10^{6}$ & $8.5\%$ & $499$ & $0.2\times10^{6}$ \\
Pr$136$ & $19.7\%$ & 2958 & $1.1\times10^{6}$ & $19.1\%$ & 623 & $0.5\times10^{6}$ & $13.0\%$ & $1284$ & $0.1\times10^{6}$ \\
Ts$225$ & $22.2\%$ & 8206 & $3.5\times10^{6}$ & $29.0\%$ & 1037 & $0.6\times10^{6}$ & $14.7\%$ & $1901$ & $0.2\times10^{6}$ \\
att$532$ & $43.0\%$ & $176 575$ & $4.0\times10^{6}$ & $46.0\%$ & $15376$ & $1.0\times10^{6}$ & $28.0\%$ & $20316$ & $0.2\times10^{6}$\\

\end{tabular}
\end{ruledtabular}
\label{table-1}
\end{table*}

To get a comprehensive picture of the performance of our annealing schemes, we have calculated several problems in TSPLIB$95$, for $N=48,76,101,124,136,225,532$. The annealing results in $d$-space, $\lambda$-space, $(\lambda +d)$-space are listed at Table \ref{table-1}. Besides the confirmation of the above statements about $\Delta d$, $\lambda$, and annealing velocity for $d$-space annealing and $\lambda$-space annealing, it is highly important that $(\lambda +d)$-space annealing is impressive superior to other annealing processes at each performance index.

In general, it is shown that our annealing recipe, especially $(\lambda +d)$-space annealing, is far better than CSA. But it should be stated that the results are worse and worse with the growth of the city number $N$. It is believed that the simple $3$-exchange technique is not power enough to act on TSPs with large $N$. We emphasized the fact that, in this paper, we are not to pursue a perfect algorithm for the optimal tour, but to present a distinct and excellent annealing technique.

In conclusion, for the symmetric traveling salesman problems, according to the statistical general property of the $n$th-nearest-neighbor distributions of the optimal tours, we define a new configuration space, so-called $\lambda$-space, where $\lambda$ describes the difference between the neighbor distributions of any tours with the general rule Eq. (\ref{eq02}). As indicated in simulation results, $\lambda$-space annealing is more efficient than normal simulated annealing and the combination of $\lambda$-space annealing and CSA is more powerful. Further studies would be needed to clarify more detail properties of the $n$th-nearest-neighbor distribution and to seek for more valuable recipes for any TSPs even in the case of large $N$. We expect that this new general property Eq. (\ref{eq02}) can be combined with other techniques for TSP or other optimization problems.

Note that the similar distribution in the random link TSP and the Euclidean TSP is reported in Ref. \cite{percus} after we presented this work to axriv.

\bigskip

We thanks Olivier Martin for pointing out reference \cite{percus}. The work reported in this paper was supported by the National Natural Science Foundation of China under Grant No. $10305005$ and the Special Fund for Doctor Programs in Lanzhou University.


\begin{thebibliography}{0}

\bibitem{tsp1} E. Lawler, J.K. Lenstra, R. Khan, and D. shmoys, {\it The traveling salesman problem}, (Wiley, New York, 1985).

\bibitem{tsp2} C.H. Papadimitriou and K. Steiglitz, {\it Combinatorial Optimization: Algorithms and Complexity}, (Prentice Hall, Englewood Cliffs, New Jersey, 1982).

\bibitem{tsp3} see http://www.tsp.gatech.edu/.

\bibitem{1} M. Padberg and G. Rinaldi, Oper. Res. Lett. {\bf 6}, 1(1987).

\bibitem{nn1} J.J. Hopfield and D.W. Tank, Biol. Cybernet. {\bf 52}, 141(1985).

\bibitem{sa1} S. Kirkpatrick, C.D. Gelatt, and M.P. Vecchi, Science {\bf 220}, 671(1983).

\bibitem{sa2} V. Cerny, J. Optim. Theory Appl., {\bf 45}, 41(1985).

\bibitem{hy1} N. Kawashima and M. Suzuki, J. Phys. A {\bf 25}, 1055(1992).

\bibitem{re1} T. Nagatani, Phys. Rev. A {\bf 36}, 5812(1987).

\bibitem{re2} H. Nakanishi and F. Family, Phys. Rev. A {\bf 32}, 3606(1985).

\bibitem{re3} U. Yoshiyuki and K. Yoshiki, Phys. Rev. Lett. {\bf 75}, 1683(1995).

\bibitem{diff} R. Vgajin, Physica A {\bf 307}, 260(2002).

\bibitem{gemann1984} S. Gemann and D. Gemann, IEEE Trans. Pattern Anal. Mach. Intell. {\bf PAMI-6}, 721(1984).

\bibitem{Szu1987} H. Szu and R. Hartley, Phys. Lett. A {\bf 122}, 157(1987).

\bibitem{Tsallis1994} D.A. Stariolo and C. Tsallis, in {\it Annual Review of Computational phyics II}, edited by D. Stauffer (World Scientific, Singapore, 1994).

\bibitem{Penna1995} T.J.P. Penna, Phys. Rev. E {\bf 51}, R1(1995).

\bibitem{qa1} J. Brooke, D. Bitko, and T.F. Rosenbaum, Science {\bf 284}, 779(1999).

\bibitem{qa2} R. Martonak, G.E. Santoro, and E. Tosatti, Phys. Rev. E {\bf 70}, 057701(2004).

\bibitem{tsplib} http://elib.zib.de/pub/Packages/mp-testdata/tsp/tsplib/tsplib.html.

\bibitem{lin1965} S. Lin, Bell System Tech. J. {\bf 44}, 2245(1965).

\bibitem{combine} O.C. Martin and S.W. Otto, Ann. Operat. Res. {\bf 63}, 57(1996).

\bibitem{percus} A.G. Percus and O.C. Martin, J. Stat. Phys. {\bf 94}, 739(1999).

\end{thebibliography}
\end{document}